\documentclass[aps,onecolumn,superscriptaddress]{revtex4-2}

\usepackage{graphicx,color,setspace}
\usepackage[utf8]{inputenc}
\usepackage{mathtools}
\usepackage[normalem]{ulem}
\usepackage{tabularx}
\usepackage{enumerate}%
\usepackage{float}
\usepackage[normalem]{ulem}
\usepackage{comment}
\usepackage{soul}
\usepackage{ragged2e} %
\usepackage{subcaption}
\DeclareCaptionJustification{justified}{\justifying}

\usepackage{pifont}
\let\oldding\ding% Store old \ding in \oldding
\renewcommand{\ding}[2][1]{\scalebox{#1}{\oldding{#2}}}% Scale \oldding 

\usepackage[labelfont=bf]{caption}

\begin{document}

\author{Francesco Tosti Guerra}
\affiliation{Dipartimento di Fisica, Sapienza Universit\`{a} di Roma, P.le Aldo Moro 5, 00185 Rome, Italy}

\author{Federico Marini}
\affiliation{Dipartimento di Fisica, Sapienza Universit\`{a} di Roma, P.le Aldo Moro 5, 00185 Rome, Italy}

\author{Francesco Sciortino}
\affiliation{Dipartimento di Fisica, Sapienza Universit\`{a} di Roma, P.le Aldo Moro 5, 00185 Rome, Italy}

\author{Lorenzo Rovigatti}
\email[The author to whom correspondence should be addressed, ]{lorenzo.rovigatti@uniroma1.it}
\affiliation{Dipartimento di Fisica, Sapienza Universit\`{a} di Roma, P.le Aldo Moro 5, 00185 Rome, Italy}

\title{Entropy-driven phase behaviour of all-DNA associative polymers}

\begin{abstract}
Associative polymers (APs) with reversible, specific interactions between ``sticker'' sites exhibit a phase behavior that depends on a delicate balance between distinct contributions controlling the binding. For highly-bonded systems, it is entropy that mostly determines if, on increasing concentration, the network forms progressively or \textit{via} a first-order transition.
With the aim of introducing an experimentally-viable system tailored to test the subtle dependence of the phase behavior on the binding site topology,  here we numerically investigate AP polymers made of DNA, where ``sticker'' sites made by short DNA sequences are interspersed in a flexible backbone of poly-T spacers. Due to their self-complementarity, each binding sequence can associate with another identical sticky sequence.  We compare two architectures: one with a single sticker type, $(AA)_6$, and one with two distinct alternating types, $(AB)_6$. At low temperature, when most of the stickers are involved in a bond, the $(AA)_6$ system remains homogeneous, while the $(AB)_6$ system exhibits phase separation, driven primarily by entropic factors, mirroring predictions from simpler bead-spring models. Analysis of bond distributions and polymer conformations confirms that the predominantly entropic driving force behind this separation arises from the different topological constraints associated with intra- versus inter-molecular bonding. Our results establish DNA APs as a controllable, realistic platform for studying in the laboratory how the thermodynamics of associative polymer networks depends on the bonding site architecture in a clean and controlled way.
\end{abstract}

\maketitle

\section{Introduction}

Associative polymers (APs) are a class of macromolecules that contain specific interaction sites, often referred to as ``stickers'', distributed along an otherwise inert polymeric backbone~\cite{choi2020stickers,ginell_introduction_2023}. The stickers can reversibly bind to each other, giving rise to a wide range of physical behaviors, including self-assembly, phase separation, and the formation of transient networks~\cite{semenov_thermoreversible_1998, feldman2009model,seiffert2012physical}. Such dynamic and tunable properties make associative polymers highly relevant both in materials science and in the study of biological systems~\cite{seo2008polymeric, whitaker2013thermoresponsive, brangwynne2015polymer, wang2016classical, prusty2018thermodynamics}.

In recent years, associative polymers have been extensively employed as model systems to understand the behavior of a diverse range of biopolymers comprising nucleic acids and proteins that can bind through both specific and non-specific multivalent interactions, which have been shown to be able to drive phase separation and compartmentalization without membranes~\cite{nakamura2018intracellular,pappu2023phase}. To capture these complex behaviors, associative polymers are typically modeled using coarse-grained representations, where the backbone is treated as a polymer chain and the stickers as discrete sites capable of specific pairwise interactions~\cite{choi2020stickers,ginell_introduction_2023}. In the case of simple models, where chains are flexible, spacers interact through excluded volume only, and each sticker can be involved in no more than one bond, there is theoretical~\cite{semenov_thermoreversible_1998}, numerical~\cite{formanek2021gel,paciolla2021validity,chen_solgel_2025}, and experimental~\cite{whitaker2013thermoresponsive,tang2015anomalous} evidence that no thermodynamic instability is encountered upon increasing concentration. Indeed, in the dilute regime most of the bonds involve stickers belonging to the same chain (intra-molecular bonds), which therefore resembles a single-chain nanoparticle~\cite{pomposo2017single}. However, as concentration increases, chains start to interact more strongly with each other, forming inter-molecular connections that are entropically favoured~\cite{sciortino2019entropy,sciortino2020combinatorial}, and therefore progressively replace the intra-molecular bonds, thereby giving raise to a network-like structure without any phase separation occurring. However, it has been recently shown that in simple bead-spring models the different entropic contributions that control the overall bonding process can be manipulated by introducing multiple sticker types, with the constraint that only like stickers can bind to each other. Interestingly, as soon as there are at least two sticker types alternating along the backbone, an entropy-driven phase separation appears for a wide range of system parameters~\cite{rovigatti2022designing, rovigatti2023entropy}, akin to other soft-matter phase transitions induced by entropy, such as those due to depletion~\cite{asakura1954interaction,tuinier2011colloids}, combinatorial~\cite{zilman2003entropic}, and shape-driven~\cite{frenkel1982monte,lee2019entropic} interactions. The main contribution dictating the phase behaviour is due to the much larger entropic cost of closing an intra-molecular bond compared to an inter-molecular bond, effectively enhancing the number of inter-molecular contacts, and therefore the overall tendency to phase separate.

To verify that the unexpected results reported in recent numerical studies~\cite{rovigatti2022designing, rovigatti2023entropy} — based on coarse-grained polymer models — are indeed accurate, it is essential to reproduce them using a realistic polymer system that can be tested experimentally. To this end, we propose a model that, although still investigated numerically, closely mirrors a system that can be realized in the laboratory. Specifically, we represent associative polymers using a carefully designed single-stranded DNA. This is achieved by leveraging the selectivity of short base-pair binding sequences embedded within an otherwise inert poly-thymine (poly-T) strand to encode the desired binding architecture. As a result of this design, each  DNA polymer (single strand) consists of several short self-complementary sequences acting as stickers, separated by poly-thymine spacers that provide flexibility and steric separation. This architecture allows precise control over valency and interaction specificity, making it an ideal platform for probing the phase behavior and network formation of associative polymers at the molecular level. In particular, here we design DNA strands that mimic APs decorated with one or two types of stickers in order to provide an experimentally-amenable realisation of the entropic phase separation reported in Refs.~\cite{rovigatti2022designing, rovigatti2023entropy}. We find that, as for the bead-spring model, DNA APs with a single sticker type does not phase separate, whereas phase separation is observed for two-sticker-type DNA strands.  We find that also in the DNA model, despite the presence of electrostatic, excluded volume and bending energy contributions, the separation phenomenon still retains a large favourable entropic term arising from the combinatorial competition between intra- and inter-molecular bonds, as seen in the simple bead-spring model.

\section{Methods}

\begin{figure}[!ht]
\centering
\includegraphics[width=0.9\textwidth]{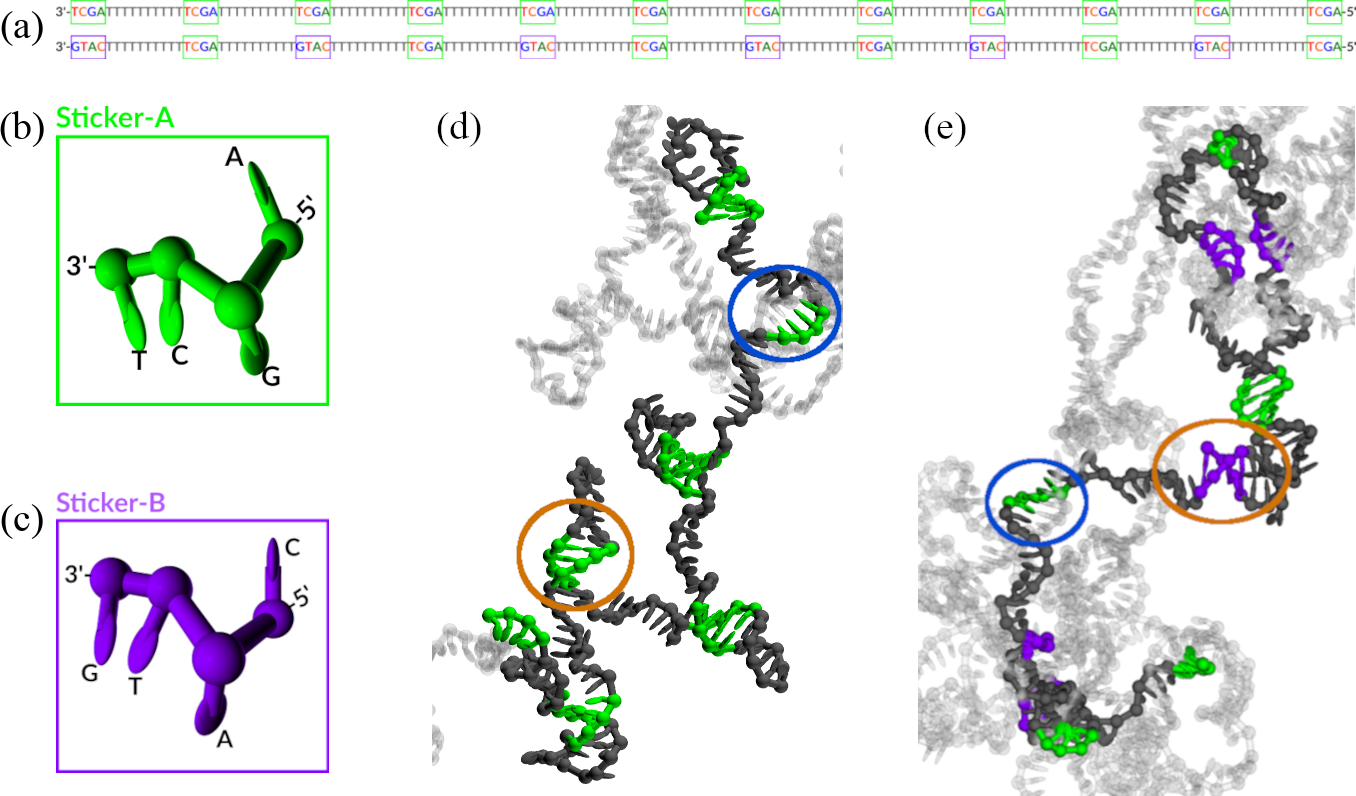}
\caption{(a) The sequences of the (top) $(AA)_6$ and (bottom) $(AB)_6$ DNA strands. (b)-(c) The sequences of the $A$ and $B$ stickers. (d) The conformation of an $(AA)_6$ strand extracted from a simulation, together with two other strands that are bonded to it, coloured in semi-transparent grey. The spacers and stickers of the main strand are depicted in dark grey and green, respectively, while the orange and blue circles highlight an intra-molecular and an inter-molecular bond, respectively. (e) Same as (d) but for an $(AB)_6$ strand; in this case there are two types of stickers, which are coloured in green and violet as in (b) and (c).}
\label{fig:cartoon}
\end{figure}

We simulate two one-component model systems made of single strand DNA in solution that we will call $(AA)_6$ and $(AB)_6$.
Each DNA strand is composed by $N_{\rm nucl} = 158$ nucleotides, with a backbone of thymine bases (the \textit{spacers}) in which are inserted $M = 12$ self-complementary regions (the \textit{stickers}) as shown in Fig.~\ref{fig:cartoon}, which also reports the full sequences of the strands used throughout this work.
 In the $(AA)_6$ systems the stickers have all the same sequence (TCGA), while in the $(AB)_6$ systems the stickers have alternating TCGA and GTAC sequences. The two sequences are self-complementary, so that only stickers of the same type can bind to each other.

The DNA strands are modelled using oxDNA~\cite{snodin2015introducing, 10.3389/fmolb.2021.693710}, and investigated by running molecular dynamics simulations with the oxDNA standalone package~\cite{rovigatti2015comparison, poppleton2023oxdna}. 
oxDNA is a coarse-grained model specifically designed to simulate the behavior of DNA at the nucleotide level. In the model each nucleotide is represented as a rigid body with effective interactions that capture essential structural, mechanical, and thermodynamic properties which makes it possible to study large-scale nucleic acid systems with high computational efficiency while maintaining physical realism. For all simulations we used the sequence-dependent version of the oxDNA2 model~\cite{snodin2015introducing}, with the salt concentration set to $S = 0.5$~M and the integration step to $\delta t = 0.002$ in reduced oxDNA units, corresponding to $\approx 6$~fs~\cite{sengar2021primer}.
In the following, we use the oxDNA length unit $\sigma = 0.8518$~nm to rescale lengths and related quantities (\textit{e.g.} density). In particular, density is expressed as the dimensionless nucleotide number density $\rho \sigma^3 = N_{\rm nucl} N_s \sigma^3 / V$, where $N_s$ is the number of strands and $V$ is the volume of the simulation box. We note for convenience that   $\rho$ can be converted to  a strand molar concentration $C_{\rm mol/l}=\rho \sigma^3 \cdot 2.69 / N_{\rm nucl}$ or to a mass concentration $C_{\rm mg/ml}=C_{\rm mol/l} \cdot M_W$, with the molecular weight ($M_W$) being $48291.6~\text{g/mol}$ for both strand types.

We run NVT simulations at several temperatures and densities starting from an initial configuration where $N_s = 128$ strands are placed randomly in the box (\textit{homogeneous simulations}). To test for the presence of a phase separation, we also perform \textit{direct-coexistence simulations} by taking an equilibrium configuration of a homogeneous simulation and elongating the box size along the $x$ direction by a factor of 4, effectively putting it in coexistence with vacuum. We then let the system evolve, monitoring the behaviour of the two initial interfaces.

To identify the gas and liquid regions in the density profile, we fit 10 equilibrium profiles extracted at regular intervals along the trajectory with a smooth analytical function based on hyperbolic tangent transitions, which are well-suited for describing diffuse interfaces in finite systems. Specifically, we model the density profile as: 
\begin{equation}
\rho(x) = \frac{\rho_\mathrm{liq} + \rho_\mathrm{gas}}{2}
- \frac{\rho_\mathrm{liq} - \rho_\mathrm{gas}}{2}
\left[\tanh\left(\frac{x - x_0}{\xi}\right) - \tanh\left(\frac{x - x_1}{\xi}\right)\right]
\left[\tanh\left(\frac{x - x_2}{\xi}\right) - \tanh\left(\frac{x - x_3}{\xi}\right)\right]
\end{equation}
where $\rho_\text{liq}$ and $\rho_\text{gas}$ are the liquid and gas densities, $\xi$ is the interfacial width, and the positions $x_0 < x_1 < x_2 < x_3$ define the boundaries of the interfaces. Specifically, $x < x_0$ and $x > x_3$ are classified as gas regions, $x_1 < x < x_2$ as liquid, while the segments between $x_0$ and $x_1$, and between $x_2$ and $x_3$ correspond to the two interfaces. 
 
During all simulations we keep track of pairs of nucleotides sharing hydrogen bonds (HBs), where two nucleotides are considered bonded if their hydrogen bond interaction energy is lower than $-0.1$ in oxDNA internal units, classifying the bond as intra- or inter-molecular if the two nucleotides belong to the same strand or to different strands, respectively. We then consider two stickers as bonded if their nucleotides are involved in at least two bonds. We thus obtain the total number of stickers involved in intra- and inter-molecular bonds, $N_{\rm intra}$ and $N_{\rm inter}$, and define the fraction of stickers involved in intra- and inter-molecular as $f_{\rm intra} = N_{\rm intra} / N_{\rm max}$ and $f_{\rm inter} = N_{\rm inter} / N_{\rm max}$, where $N_{\rm max} = M N_s$ is the total number of stickers. We also define the fraction of bonded stickers as $f_{\rm tot} = (N_{\rm intra} + N_{\rm inter})/ N_{\rm max}$.

\section{Results}

\begin{figure}[h!]
    \centering
    \includegraphics[width=0.45\textwidth]{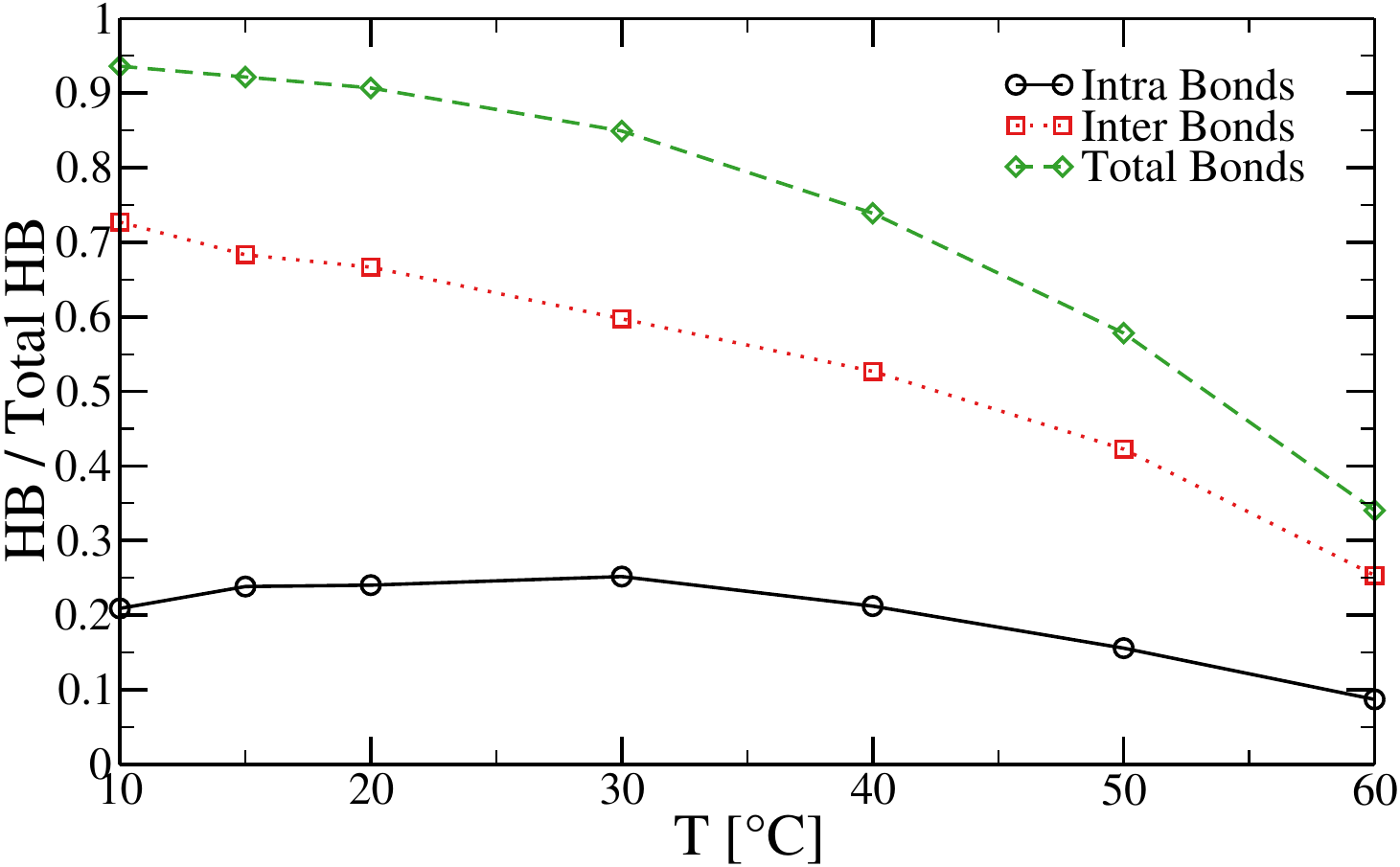}
    \includegraphics[width=0.45\textwidth]{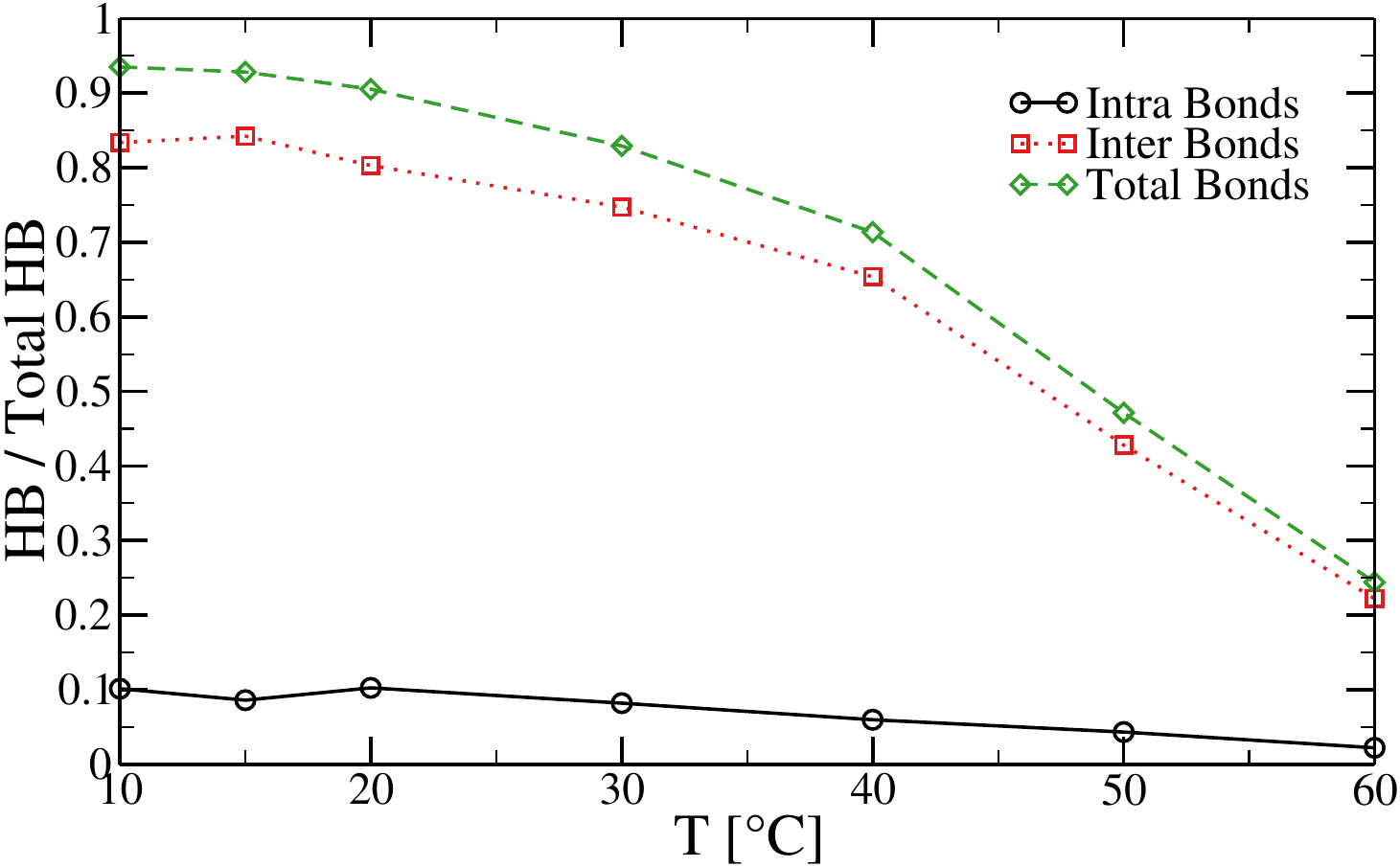}\\
    \includegraphics[width=0.45\textwidth]{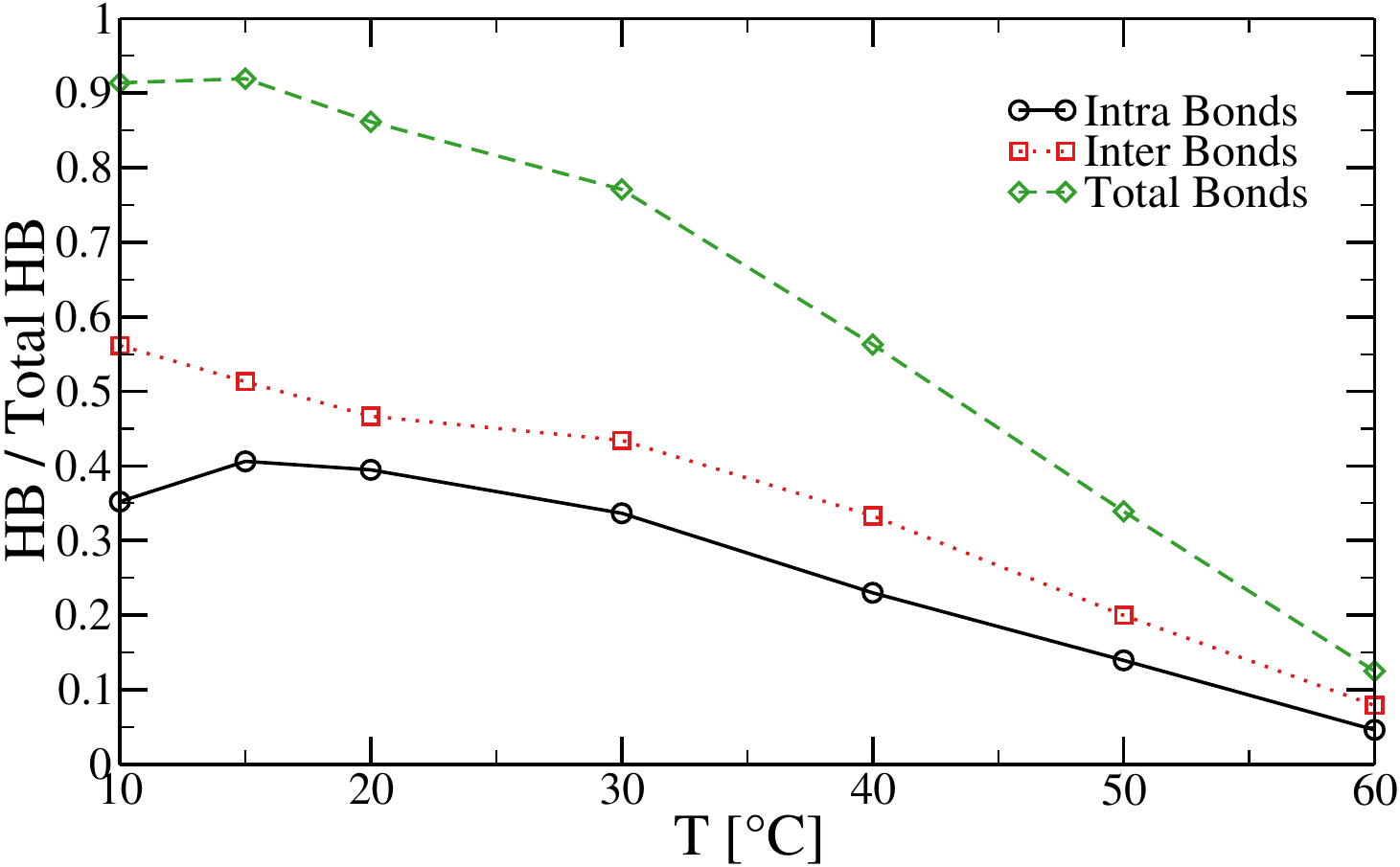}
    \includegraphics[width=0.45\textwidth]{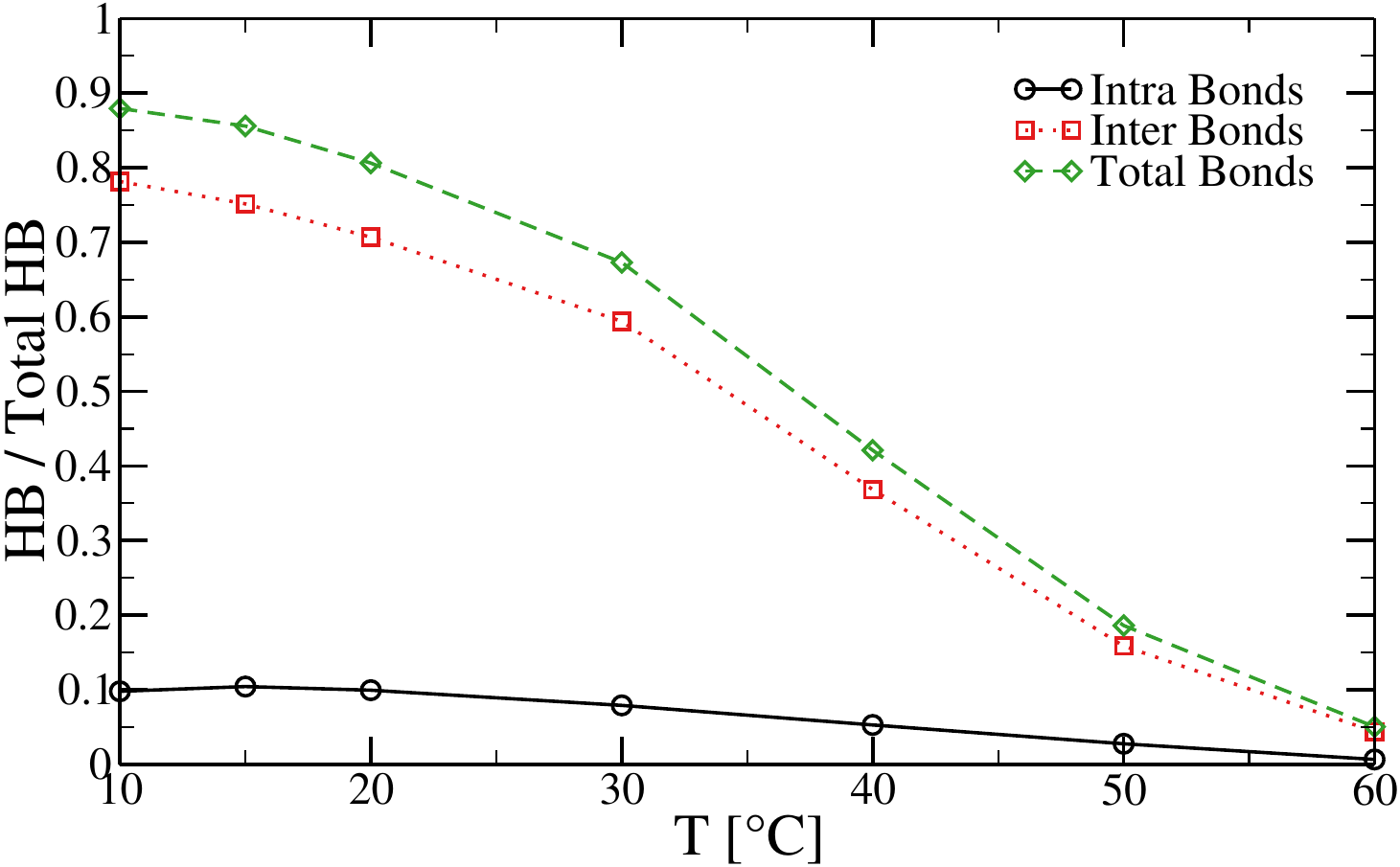}\\
    \includegraphics[width=0.45\textwidth]{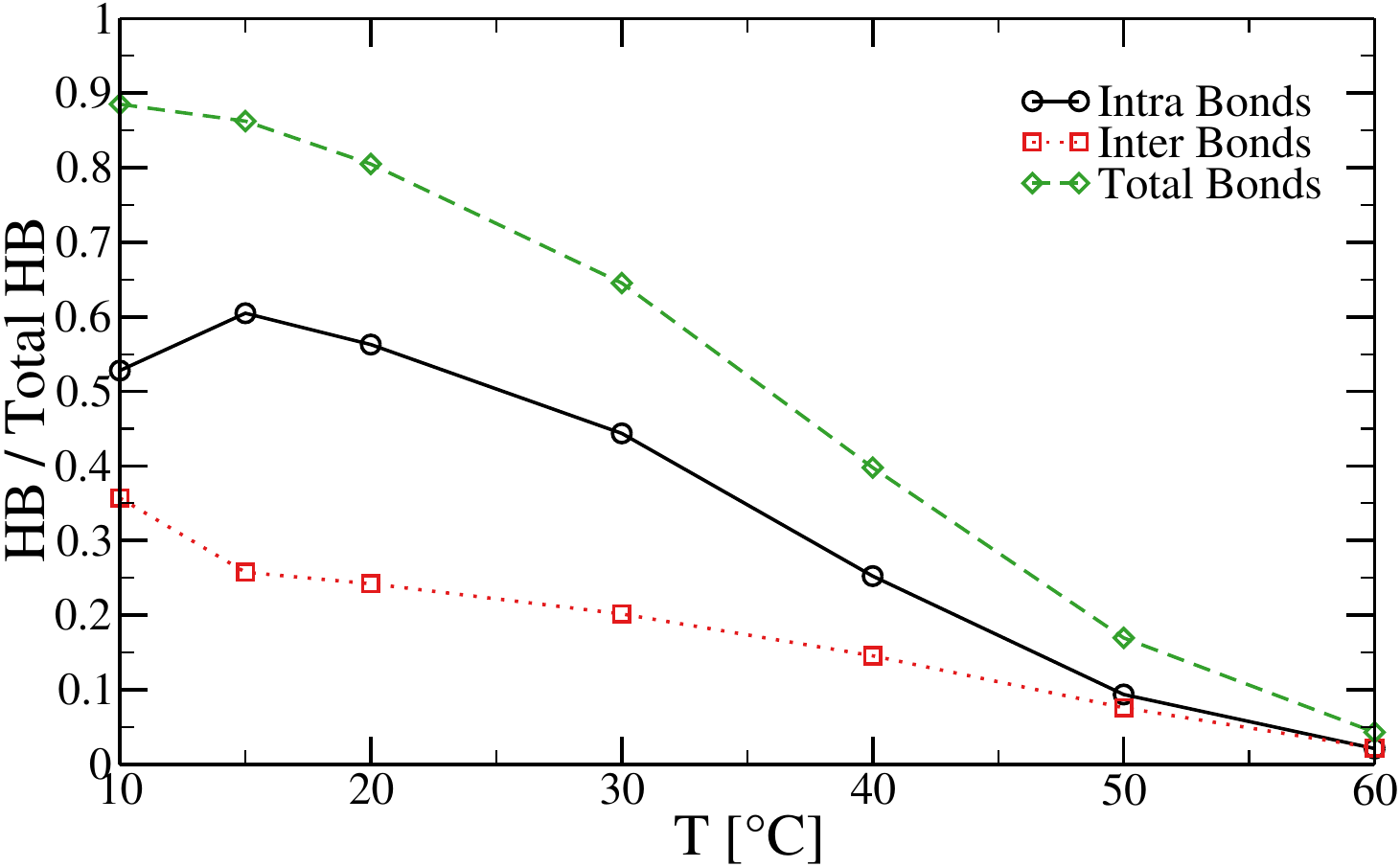}
    \includegraphics[width=0.45\textwidth]{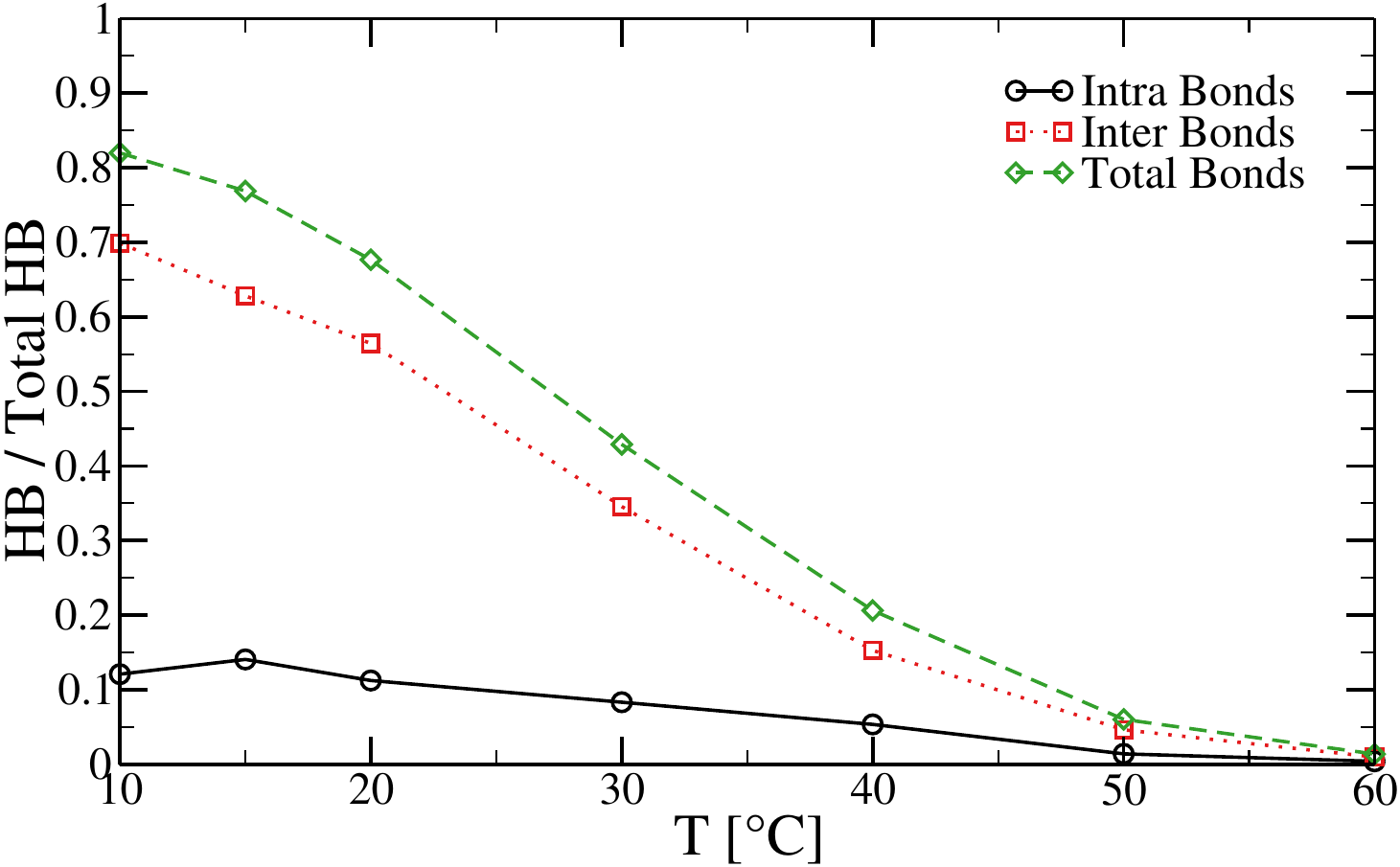}
    \caption{The fraction of hydrogen bonds (HBs) between nucleotides belonging to stickers involved in total, intra- or inter-molecular bonds for the (left) $(AA)_6$ and (right) $(AB)_6$ systems. From top to bottom, rows correspond to densities $\rho \sigma^3 = 0.57$, $\rho \sigma^3 = 0.31$, and $\rho \sigma^3 = 0.16$.}
    \label{fig:melting}
\end{figure}

We start by characterizing the bonding process in the two systems as a function of temperature and density. Figure~\ref{fig:melting} shows the temperature-dependence of the fractions of hydrogen-bonded nucleotides belonging to stickers of the $(AA)_6$ and $(AB)_6$ homogeneous systems, classified according to the type of bond (inter- or intra-molecular), as well as the sum of the two, for three different densities.

These melting curves exhibit the well-known sigmoidal shape typical of DNA pairing, characterized by a density dependent melting temperature which decreases as the density decreases.  In almost all simulations, the fraction of bonded nucleotides reaches 90\%, signaling the extensive bonding process that takes place in both systems.  Consistent with previous results derived for a bead-spring model~\cite{rovigatti2022designing,rovigatti2023entropy}, we observe a striking difference in the fraction of intra-strand bonds between the $(AA)_6$ and $(AB)_6$ systems.  In the $(AB)_6$ case the majority of the bonds are formed between sticky sequences of different strands, and the fraction of intra-strand bonds never exceeds 15\%. Differently, in the case of the $(AA)_6$ system, the fraction of intra bonds is enhanced, becoming dominating at the lowest densities. 

We also note on passing that the fraction of intra-molecular bonds nearly always displays a maximum at some intermediate temperature. We posit that this non-monotonic behaviour is due to the finite flexibility of the strands: indeed, as temperature decreases, nucleotides in the spacer regions tend to stack more, making the strands less flexible, thereby effectively hindering the formation of intra-molecular bonds.

\begin{figure}[h!]
\centering
    \includegraphics[width=0.5\textwidth]{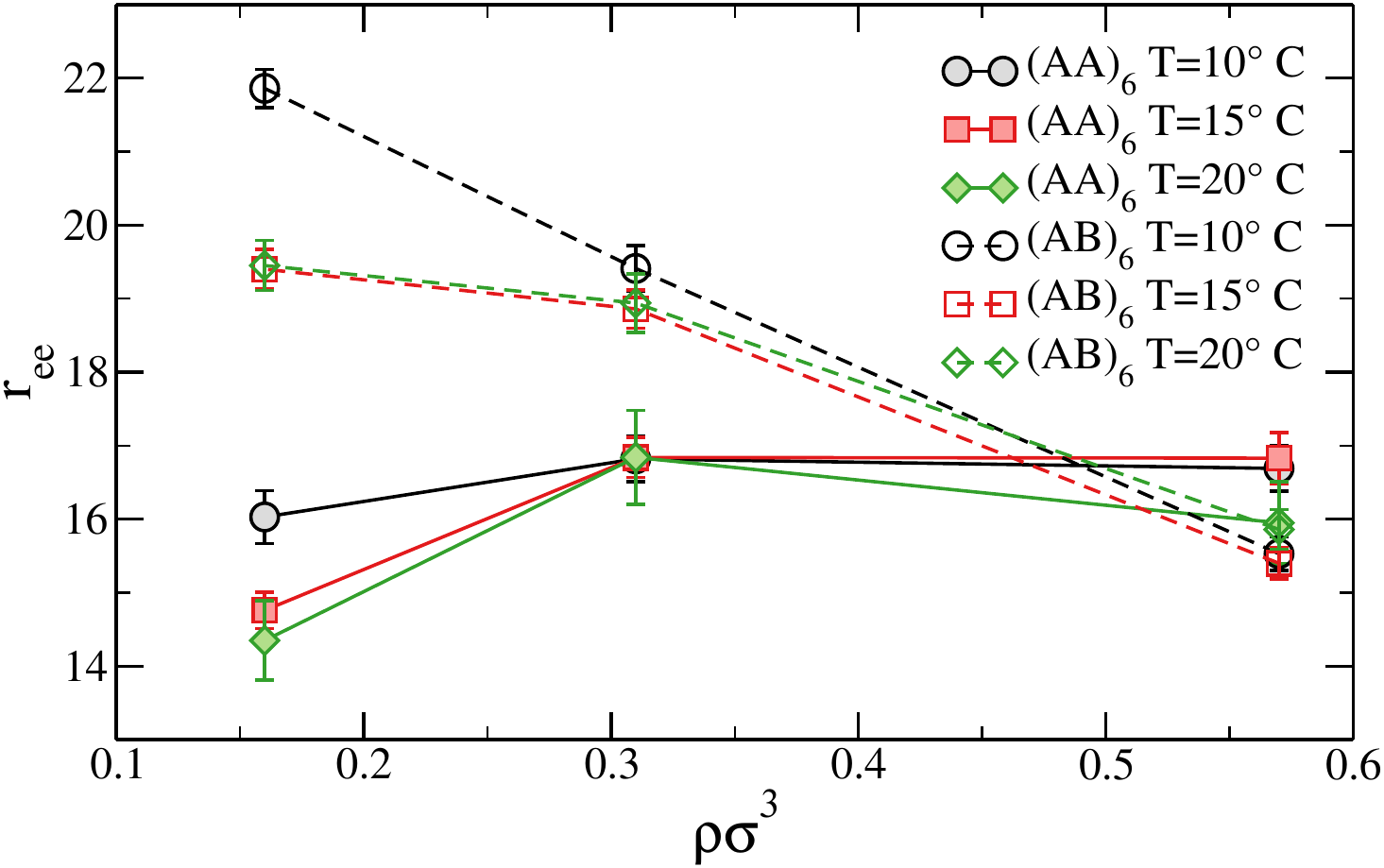}
    \caption{The average end-to-end distance of homogeneous (AA)$_6$ (full symbols joined by solid lines) and (AB)$_6$ (empty symbols joined by dashed lines) systems as a function of $\rho$, for the three lowest temperatures simulated. The error bars have been estimated by dividing the standard deviation of the mean by $\sqrt{N_s} = \sqrt{128}$, which provides an upper bound for the error.\label{fig:ree_homogeneous}
    }
\end{figure}

In order to understand how the interplay between total, inter- and intra-molecular bonds affect the conformation of the strands, we report in Figure~\ref{fig:ree_homogeneous} the average end-to-end distance of the low-temperature systems.

For the (AA)$_6$ systems, the end-to-end distance $r_{\mathrm{ee}}$ displays minimal dependence on temperature and only a weak dependence on density. Specifically, $r_{\mathrm{ee}}$ increases slightly with increasing $\rho$, which we attribute to a gradual shift from intra- to inter-molecular bonding as the system becomes denser.

The behavior of the (AB)$_6$ systems is somewhat more nuanced. While temperature again has a weak effect, $r_{\mathrm{ee}}$ decreases significantly with increasing $\rho$. This trend is likely driven by the growing total number of bonds as $\rho$ increases (see Figure~\ref{fig:melting}), which leads to a more compact molecular conformation.

Notably, at the highest density considered, the end-to-end distance of the (AA)$_6$ systems exceeds that of the (AB)$_6$ systems. As shown in Figure~\ref{fig:melting}, the total number of bonds is nearly identical for both strand types at this density. Therefore, the observed difference in $r_{\mathrm{ee}}$ can be attributed to the nature of the intra-molecular bonding: in (AB)$_6$ systems, these bonds result in more compact configurations, indicating a higher entropic cost compared to their (AA)$_6$ counterparts.

To answer the question whether the difference in sequences has indeed a relevant role in controlling the phase behavior of associative polymers in the limit of extensive bonding, we perform direct-coexistence simulations for temperatures lower than $\approx 20^\circ$~C, starting from initial configurations based on the equilibrated homogeneous samples as detailed in the methods section.

\begin{figure}
    \centering
    \begin{subfigure}[b]{0.065\textwidth}
        \includegraphics[width=\linewidth]{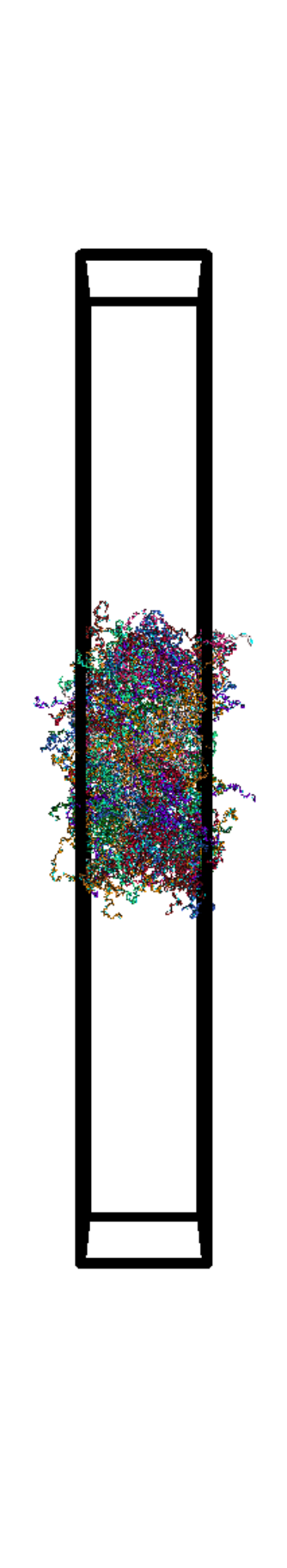}
        \caption{}
        \label{fig:init_conf}
    \end{subfigure}
    \begin{subfigure}[b]{0.3266\textwidth}
        \includegraphics[width=\linewidth]{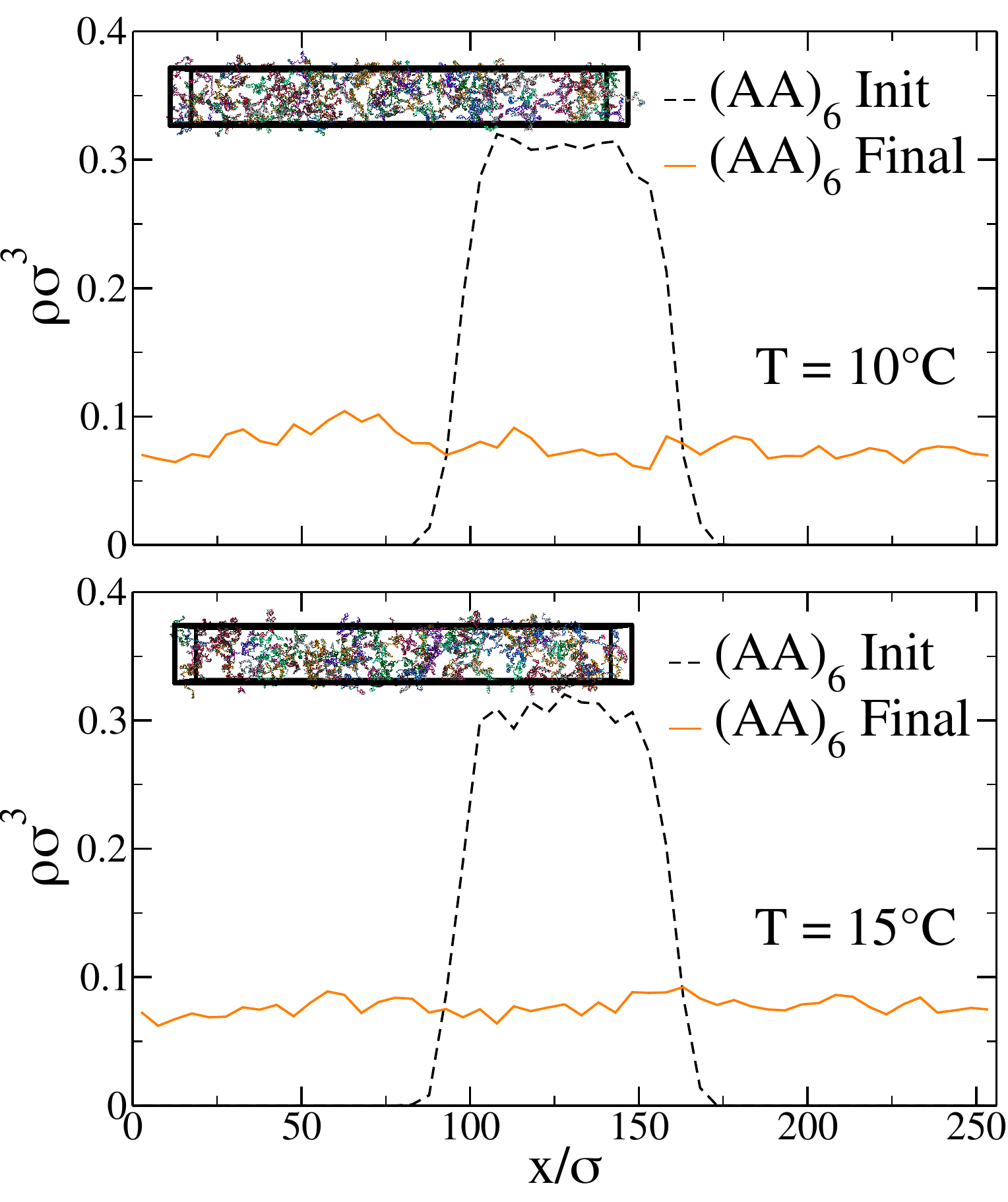}
        \caption{}
        \label{fig:dens_prof_AA}
        %\vspace{3em}
    \end{subfigure}
    \begin{subfigure}[b]{0.58\textwidth}
        \includegraphics[width=\linewidth]{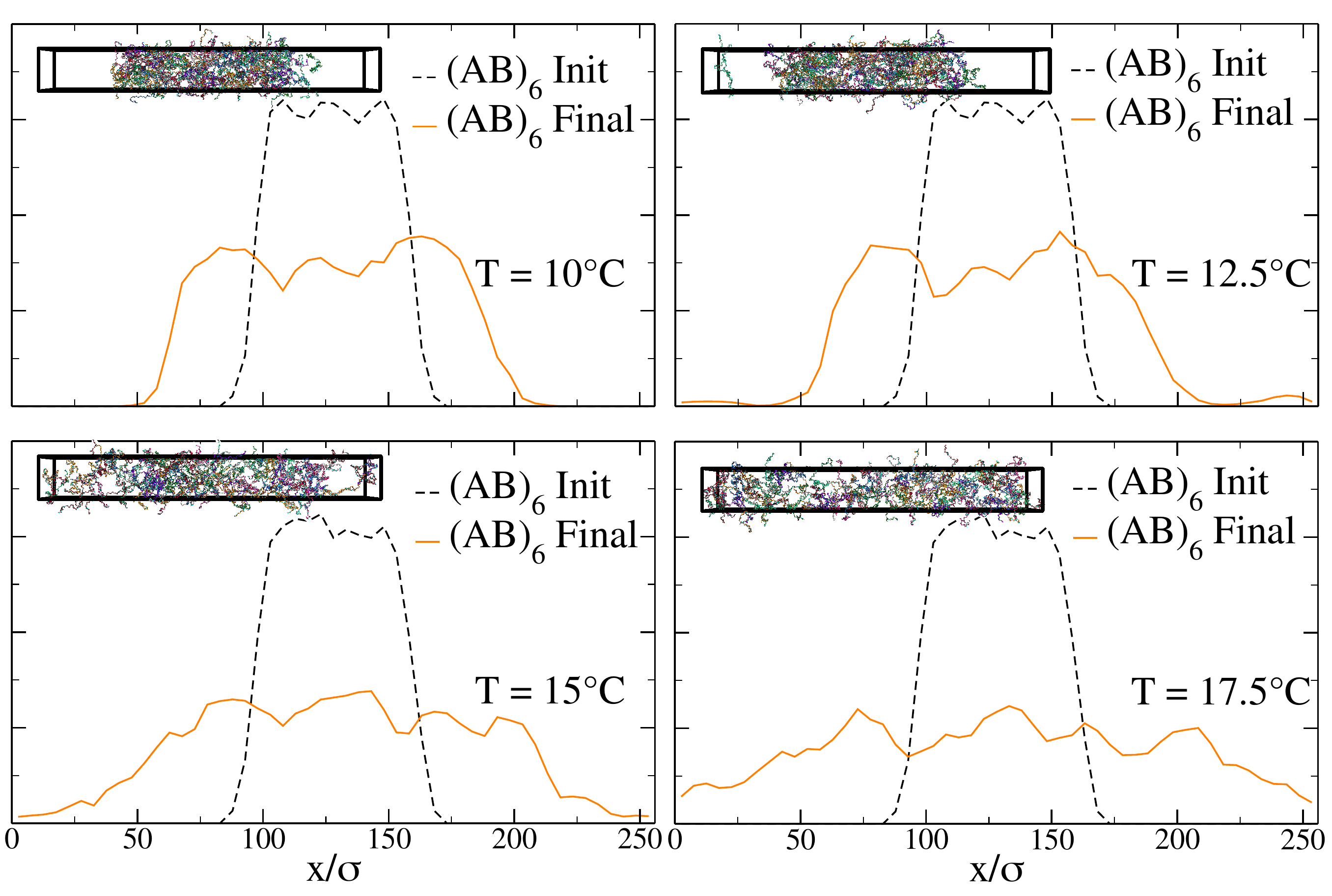}
        \caption{}
        \label{fig:dens_prof_AB}
        %\vspace{3em}
    \end{subfigure}
    \caption{(\subref*{fig:init_conf}) Graphic representation of the system in a direct-coexistence simulation at the beginning of the calculation. (\subref*{fig:dens_prof_AA}) Density profiles at the start of the simulation (dashed black curve) and at the end of the simulation, when the system has reached equilibrium (solid orange),  for the $(AA)_6$ system at two different temperatures. (\subref*{fig:dens_prof_AB}) Analogous results for the $(AB)_6$ systems for four different temperatures. In panels~(\subref*{fig:dens_prof_AA}) and~(\subref*{fig:dens_prof_AB}) characteristic snapshots of the final configurations are also shown as insets. All the direct-coexistence simulations shown here have been initialised from homogeneous starting configurations with $\rho \sigma^3 = 0.57$.}
    \label{fig:dens_prof}
\end{figure}

\begin{figure}
    \centering
    \includegraphics[width=0.45\linewidth]{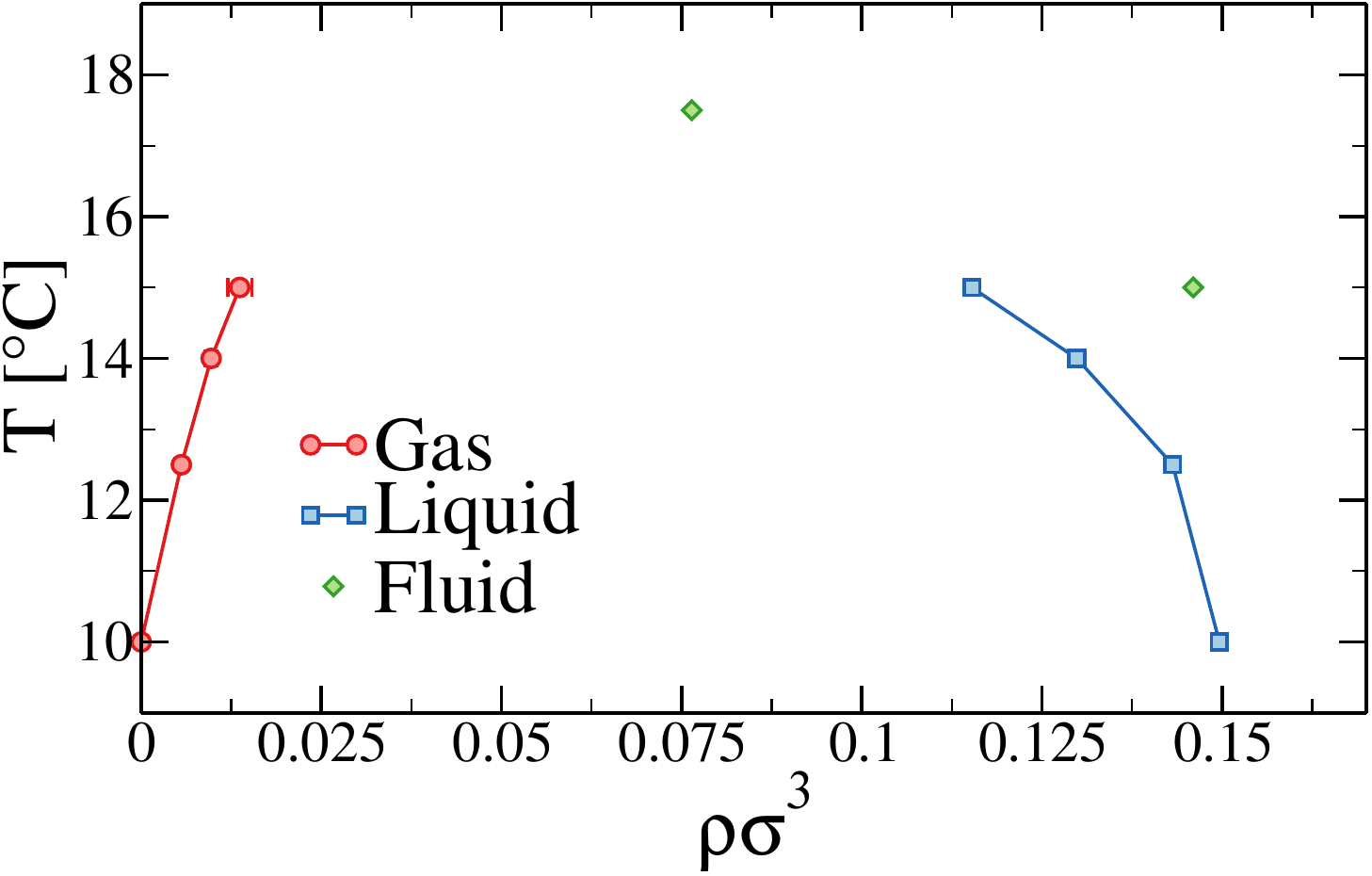}
    \caption{The phase diagram of the $(AB)_6$ system, displaying the density of the fluid, gas and liquid phases (\textit{i.e.} state points where phase separation does not occur) as extracted from direct-coexistence simulations. Note that the density of the gas at the lowest temperature ($10^\circ$~C) is approximated to zero, since no strand detached from the interfaces throughout the simulation. Note that the names ``gas phase`` and ``liquid phase'' are the "implicit solvent" analog of polymer poor and polymer rich (network) phases.}
    \label{fig:phase_diagram}
\end{figure}

As shown in the Appendix, the characteristic bond-bond autocorrelation time is $\approx 10^9$ time steps at $15^\circ$~C, and therefore we run simulations that spans longer time frames (up to $\approx 3 \times 10^{10}$ simulation steps, corresponding to 2-3 months of simulation time on a NVIDIA A100 GPU). We use the fraction of inter- and intra-molecular bonds as a proxy for equilibration, which we assume to be attained once these fractions reach a plateau value. We then compute density profiles along the elongated box, and classify each system as phase separated or homogeneous depending on whether the equilibrated profiles show the presence of interfaces or not. 

Figure~\ref{fig:dens_prof_AA} shows the initial and final (equilibrium) density profiles for the $(AA)_6$ system, for which the direct coexistence simulations evolve toward an homogeneous configuration, providing strong evidence of the absence of phase separation in favor of a continuous cross-over from the dilute to the network state on increasing density. By contrast, Figure~\ref{fig:dens_prof_AB} shows that in the $(AB)_6$ case and at low temperature the density of the dense phase relaxes progressively to an equilibrium value, but in this case retaining a clear interface with a lower-density phase, indicating that  a dilute solution of intra-bonded strands coexists with a network of mostly inter-bonded strands. It is only at $T = 17.5^\circ$~C  and above the $(AB)_6$ system approaches a homogeneous state for all densities.

From the value of the density in the dense and in the dilute phases at equilibrium it is possible to estimate the two coexisting densities. To do so in the phase-separated samples, we classify DNA strands as belonging to the interface, gas or liquid phase, making it possible to draw a phase diagram of the system, shown in Figure~\ref{fig:phase_diagram}, which also shows studied state points where the interfaces melted and the density became homogeneous during the course of the direct-coexistence simulation (green points labelled as ``fluid''). As the temperature increases, the density of the coexisting phases approach each other, and from the absence of interfaces  at $T = 17.5^\circ$~C we can predict  the critical temperature of the $(AB)_6$  system to lie between $15^\circ$~C and $17.5^\circ$~C. By evaluating the average between the two coexisting densities at $T = 15^\circ$~C we can also provide a rough estimate for the critical density, $\rho_c\sigma^3 = 0.06$,  corresponding to a strand concentration of $\approx 1$~mM or a weight density of $\approx 53$~mg/ml. This value is of the same order of magnitude of the overlap concentration $\rho^* \sigma^3 = N_{\rm nucl} / R_g^3 \approx 0.18$ (\textit{i.e.} $\approx 3$~mM strand concentration), as estimated by using the data of the coexisting liquid phase at $T = 15^\circ$~C (see below for a more thorough discussion), for which we find $R_g \approx 9.5\, \sigma$. This is in line with expectations, since in associative polymer systems chain-chain interactions becomes important around the overlap concentration~[3].

\begin{figure}[ht]
    \centering
    \begin{subfigure}[b]{0.32\textwidth}
        \includegraphics[width=\textwidth]{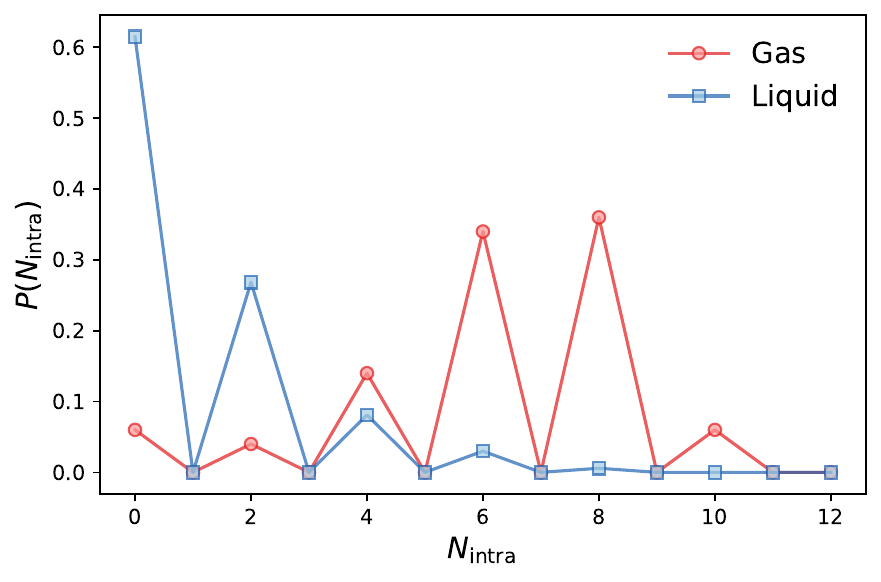}
        \caption{}
    \end{subfigure}
    \hfill
    \begin{subfigure}[b]{0.32\textwidth}
        \includegraphics[width=\textwidth]{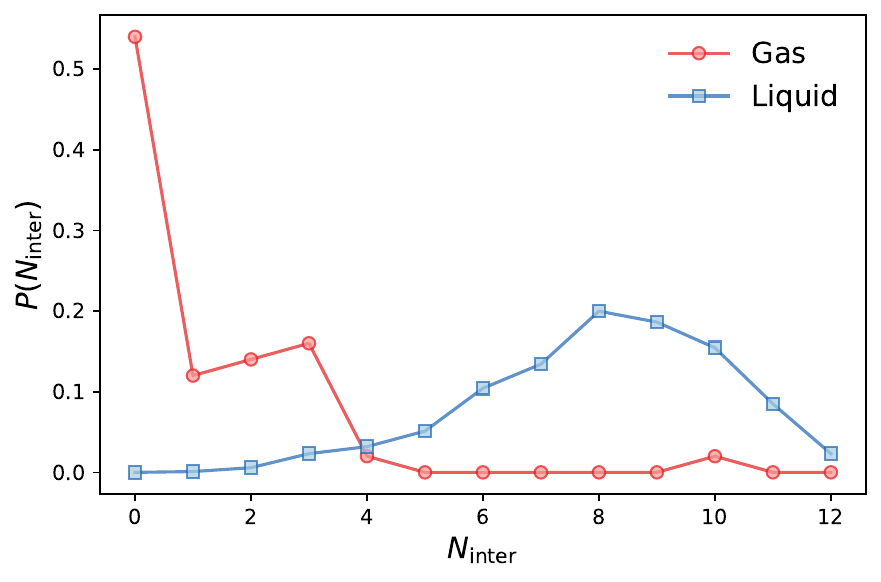}
        \caption{}
    \end{subfigure}
    \hfill
    \begin{subfigure}[b]{0.32\textwidth}
        \includegraphics[width=\textwidth]{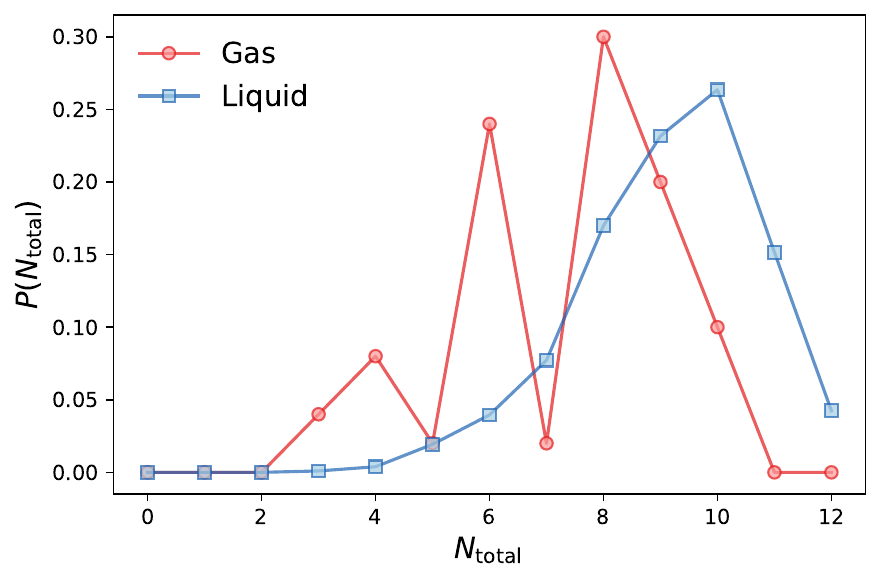}
        \caption{}
    \end{subfigure}
    
    \vspace{0.5cm} 
    \begin{subfigure}[b]{0.32\textwidth}
        \includegraphics[width=\textwidth]{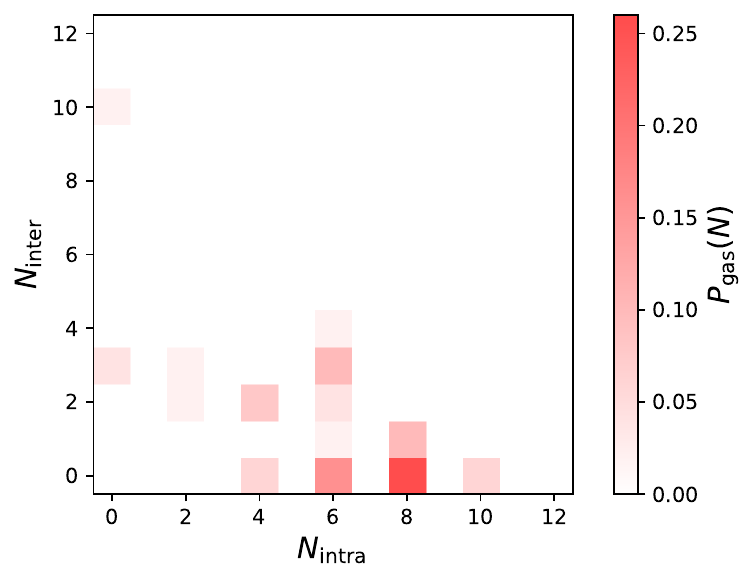}
        \caption{}
    \end{subfigure}
    \begin{subfigure}[b]{0.32\textwidth}
        \includegraphics[width=\textwidth]{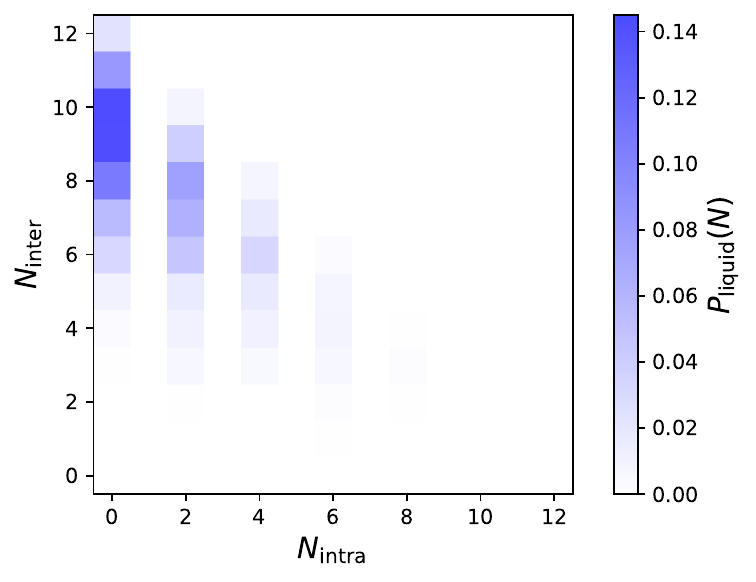}
        \caption{}
    \end{subfigure}

    \caption{Normalised histograms of the bonds found in the phase-separating $(AB)_6$ system simulated at $T = 15^\circ$~C for strands belonging to the gas and liquid phases (see text for definitions). Panels (a), (b) and (c) show the one-dimensional histograms for the intra-molecular, inter-molecular, and total bonds, respectively. Panels (d) and (e) show the two-dimensional histograms for the intra- and inter-molecular bonds of the strands found in the gas and liquid phases, respectively.}
    \label{fig:bond_histograms}
\end{figure}

As shown in Fig.~\ref{fig:phase_diagram}, for $T = 15^\circ$~C the fraction of DNA strands in the gas is not negligible. This makes it possible to compare the properties of the single strands that make up the two phases for this particular state point. In particular, we investigate the difference in binding properties of the two phases. While the current number of gas-phase strands provides only preliminary statistical insights, we aim to enhance these results in future work by performing targeted simulations of the two phases.

Figure~\ref{fig:bond_histograms} shows the histograms of the inter-strand, intra-strand and total sticky sequences involved in bonds.  The zig-zag pattern in the intra-strand histogram originates from the fact that each intra-strand bond involves two sticky sequences.  
 As expected, the low density phase is rich in intra-molecular bonds and poor in inter-molecular bonds, while the reverse is true for the high density phase. At this temperature, which is the only one for which we have a sufficient number of gas strands to analyse, the two phases slightly differ in the total number of bonds ($\approx 9$ and $\approx 7.5$ average bonded stickers for the liquid and gas strands, respectively), signaling a weak energetic contribution to the difference in chemical potential. Since, as shown in Fig.~\ref{fig:melting}, the fraction of formed bonds increases upon decreasing temperature regardless of the density, we expect that the observed small difference in the number of bonds will vanish at temperatures lower than those investigated here, strongly suggesting that the two phases will have the same energy and thus that the separation is indeed driven (especially at low temperature) by a strong entropic component stemming from the different conformations and different topology of the bonds adopted by the strands in the two phases.

 %Although the average fraction of bonded stickers is $\approx 0.75 $ in the high density and $\approx 0.6 $ in the low density.   corresponding to $\approx 9$ and $\approx 7.5$ average bonded stickers
 
\begin{figure}[ht]
    \centering
    \begin{subfigure}[b]{0.45\textwidth}
        \includegraphics[width=\textwidth]{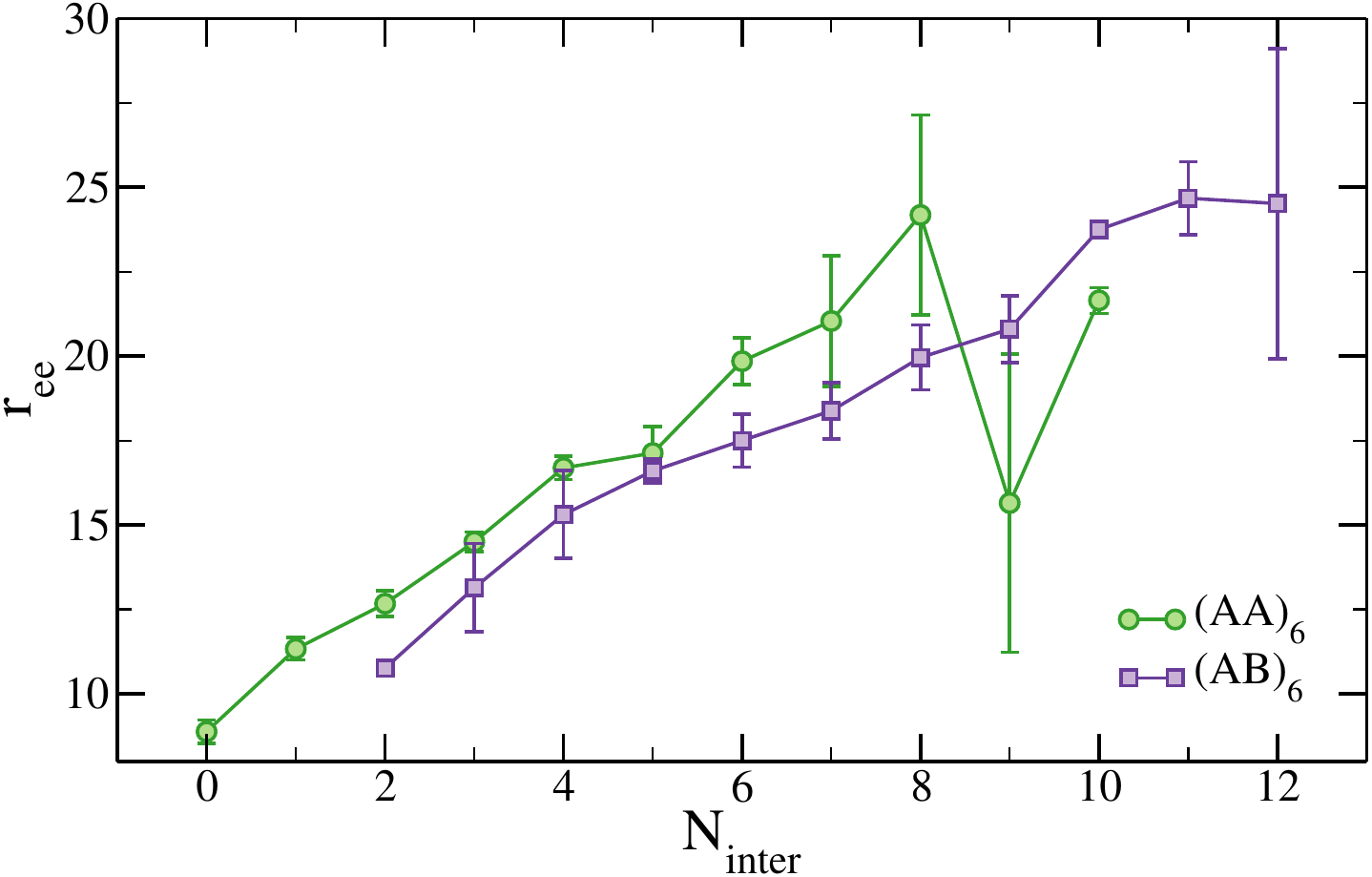}
        \caption{}
    \end{subfigure}
    \begin{subfigure}[b]{0.45\textwidth}
        \includegraphics[width=\textwidth]{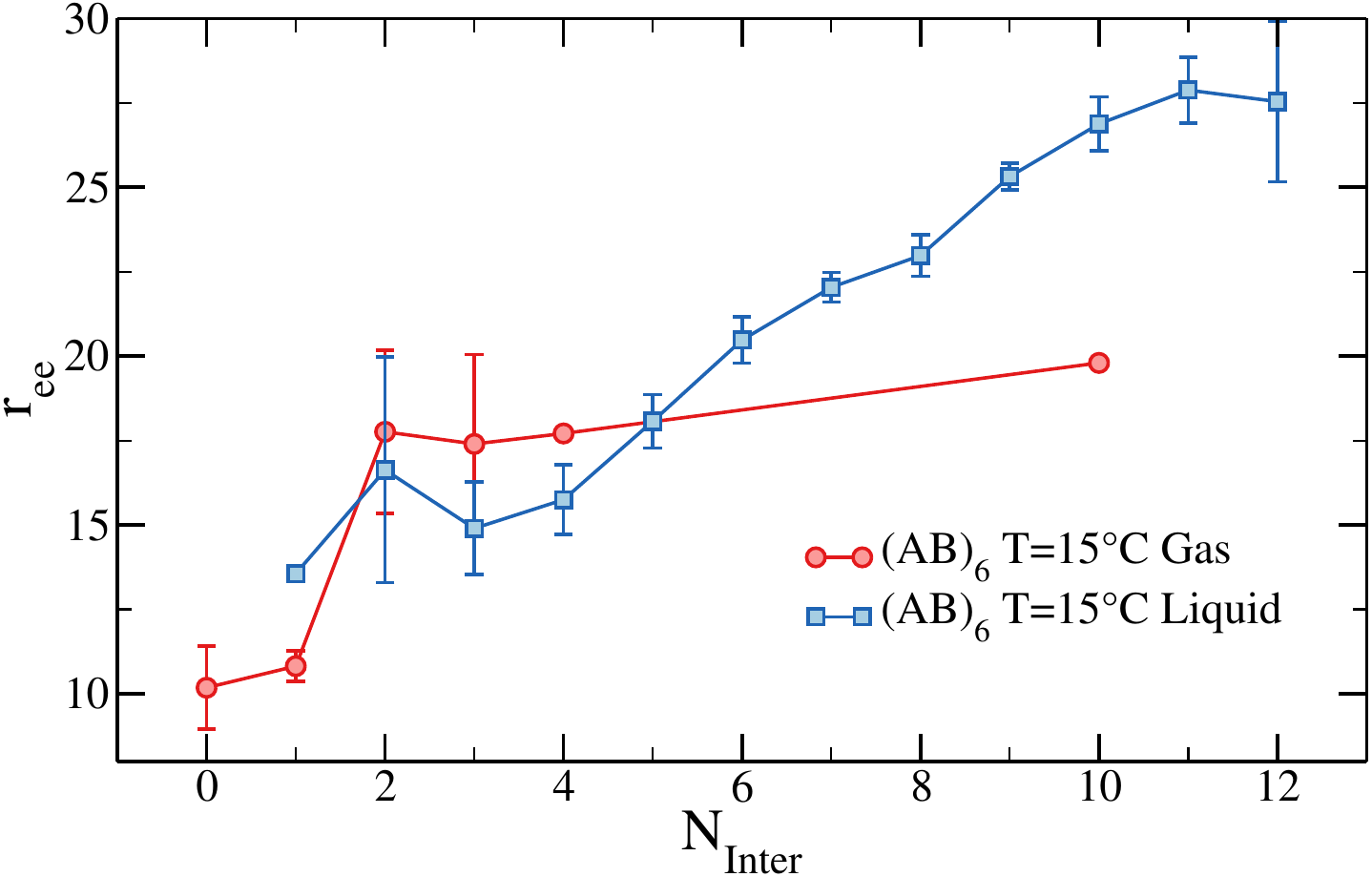}
        \caption{}
    \end{subfigure}
        \caption{(a) The average end-to-end distance, $r_{ee}$ as a function of the number of stickers involved in inter-molecular bonds, $N_{\rm inter}$, for the $(AA)_6$ and $(AB)_6$ homogeneous systems simulated at $T = 15^\circ$~C and $\rho\sigma^3 = 0.16$. (b) $r_{ee}$ as a function of $N_{\rm inter}$ for the gas and liquid phases found in the direct-coexistence $(AB)_6$ system simulated at $T = 15^\circ$~C. In both panels error bars show the mean-squared error, and therefore are absent for points obtained from single measurements (see \textit{e.g.} the $N_{\rm inter} = 10$ point in (b)).}
    \label{fig:Ree}
    \end{figure}
    
To support this claim, we investigate the conformation of the strands extracted from the homogeneous simulations. Figure~\ref{fig:Ree}(a) shows a comparison between the average end-to-end distance of $(AA)_6$ and $(AB)_6$ strands, $r_{ee}$, as a function of the number of stickers involved in inter-molecular bonds. Since both the systems investigated are almost fully bonded, more inter-molecular bonds means fewer intra-molecular bonds, which tend to compact the strands. As a result, we observe that $r_{ee}$ increases monotonically with $N_{\rm inter}$. The roughly linear behaviour can be explained by assuming that each additional broken intra-molecular bond contributes to the end-to-end distance by the same amount, since each bond constraints portion of the strands that are approximately the same length. As expected, $(AB)_6$ strands are always smaller than $(AA)_6$ strands, since each intra-molecular bond closes a loop that is at least twice in size that of an $(AA)_6$ strand. It is this entropic penalty that shifts the balance towards inter-molecular bonds, in line with previous results on bead-spring models~\cite{rovigatti2022designing}.

If we now focus on the phase-separated $(AB)_6$ system simulated at $T = 15^\circ$~C, we see in Figure~\ref{fig:Ree}(b) that, as expected, in the gas the small number of inter-molecular bonds leads to compact conformations, whereas the liquid, with its higher inter-molecular connectivity, features larger end-to-end distances. Such a large difference in the conformational freedom of the strands tips the balance towards phase separation in the $(AB)_6$ system, providing a realistic example of the fully-entropic phase separation observed in bead-spring models.

\section{Conclusions}

In this work, we have presented a computational study of DNA-based associative polymers, designed to emulate the behavior of idealized APs, while at the same time providing an experimentally realisable system. 
The capacity to design sticky-end DNA sequences that are simultaneously self-complementary and mutually orthogonal makes this fully DNA-based associative polymer a versatile and powerful platform for encoding diverse bond architectures.

By leveraging the oxDNA model, we simulated single-stranded DNA constructs featuring either one or two types of self-complementary sticky sequences, separated by flexible poly-T spacers. Our findings show that at low temperatures on increasing concentration, systems with a single sticker type progressively form highly bonded, homogeneous networks, while those with alternating sticker types   undergo a sharp first order phase separation, nucleating a connected network of polymers. 

This phase separation arises from an entropic driving force associated with the suppression of intra-molecular bonding in favor of more flexible inter-molecular connections. The presence of phase separation in strands decorated with two types of stickers mirrors predictions from coarse-grained bead-spring models~\cite{rovigatti2022designing} and demonstrates that similar principles can be realized in realistic, sequence-programmable systems.

The results here presented not only validate previous work but also highlight the utility of DNA as a tunable platform for investigating the thermodynamics of associative networks. Our approach opens the door to the experimental realization of entropy-driven phase behavior in all-DNA systems, with potential implications for the design of responsive materials and hopefully for a deeper understanding of systems where phase separation of biomacromolecules is considered to be important~\cite{wang2016classical, nguyen2022condensates}.

\section*{Acknowledgements}

We acknowledge support by ICSC – Centro Nazionale di Ricerca in High Performance Computing, Big Data and Quantum Computing, funded by European Union - NextGenerationEU, and CINECA-ISCRA for HPC resources. L.R acknowledges support from MUR-PRIN Grants No. 20225NPY8P and P2022JZEJR, funded by European Union - NextGenerationEU, Missione 4 Componente 2 - CUP B53D23028570001.

\section*{Data availability}

The data that support the findings of this study are available from the corresponding author upon reasonable request.

\appendix

\section{The bond-bond autocorrelation function}

\begin{figure}[h!]
\centering
    \includegraphics[width=0.5\textwidth]{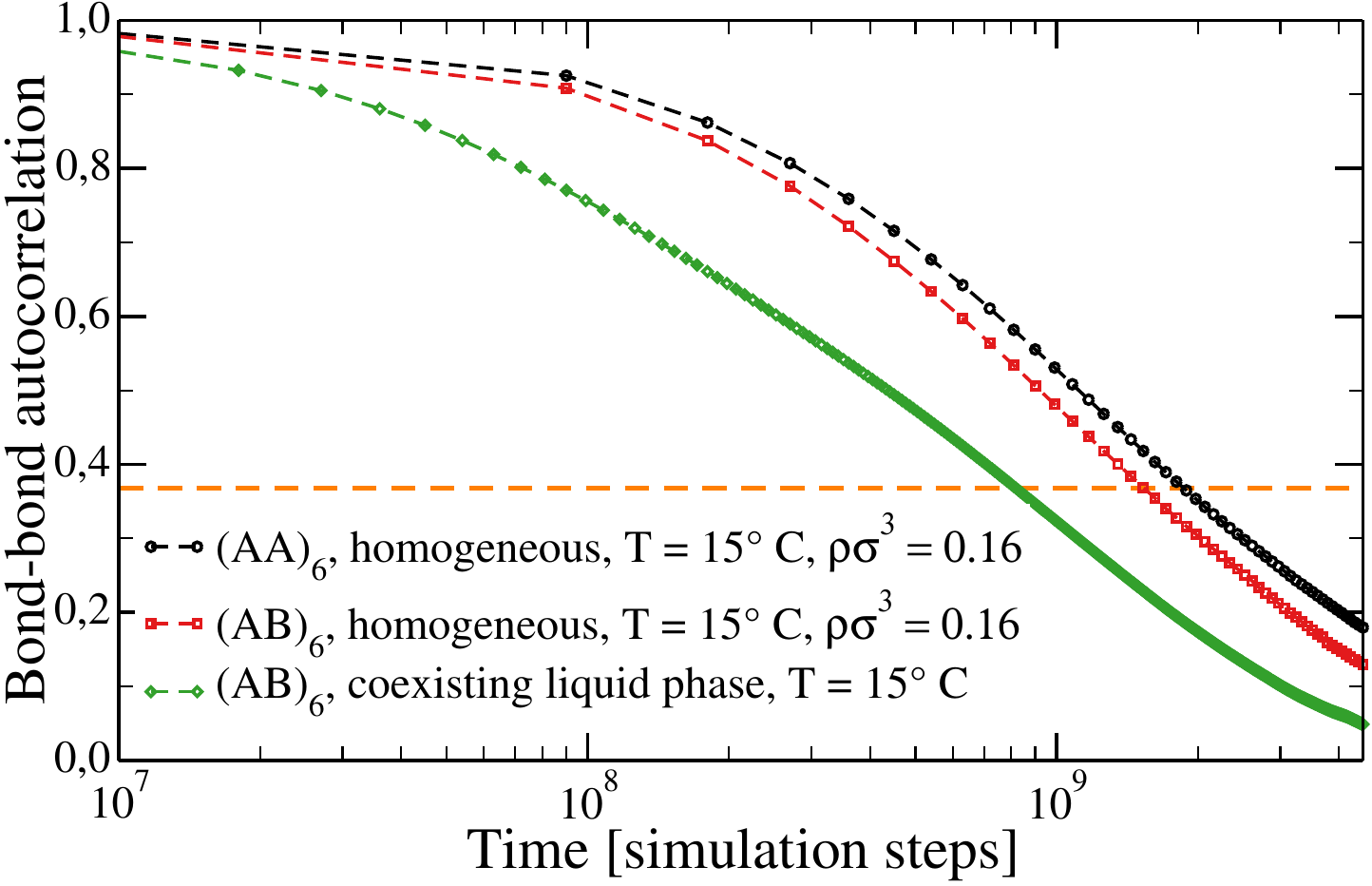}
    \caption{The bond-bond autocorrelation function for the (AA)$_6$ and the (AB)$_6$  systems in an homogeneous state point (for densities listed in the label), and  for the liquid component of the (AB)$_6$ in the direct-coexistence simulation.  All data refer to $T = 15^\circ$~C.
    The crossing between the data and the horizontal orange dashed line (drawn at a value $1 / e$) is the conventional estimate for the characteristic decorrelation time.\label{fig:bond_bond}}
\end{figure}

Figure~\ref{fig:bond_bond} shows the bond-bond autocorrelation function $B(t)$, which quantifies the fraction of bonds that were present at time $t_0$ and are still present at time $t_0 + t$, averaged over $t_0$. The figure shows that at $T = 15^\circ$~C, for which we can compare results from the homogeneous and phase-separated samples, the characteristic time scale associated to bonding, defined as the time at which $B(t) = 1/e$, is of the order of $\approx 10^9$ time steps.

\bibliography{biblio}

\end{document}